\documentclass[12pt,a4paper]{article}
\usepackage{amsmath,caption}
\usepackage[top=1.1in, bottom=1.1in, left=1.1in, right=1.1in]{geometry}
\usepackage[font={small,it}]{caption}

\usepackage[round]{natbib}
\usepackage{color,soul}

\DeclareCaptionStyle{italic}[justification=centering]{labelfont={bf},textfont={it},labelsep=colon}
\usepackage{graphicx,psfrag,epsf,textcomp}
\usepackage{enumerate,setspace}
\usepackage{natbib,dsfont}
\usepackage{url,xcolor} 
\usepackage{booktabs, subfig, bm, paralist,mathpazo,tikz,todonotes,longtable,microtype}

\usepackage[pdftex,colorlinks=true]{hyperref}
\definecolor{darkblue}{rgb}{0,0,.6}
\hypersetup{citecolor=darkblue,linkcolor=darkblue,urlcolor=darkblue}

\newcommand{\blind}{0}

\graphicspath{{plots/}}
\DeclareMathOperator*{\argmin}{\arg\!\min}
\newsavebox\CBox
\def\textBF#1{\sbox\CBox{#1}\resizebox{\wd\CBox}{\ht\CBox}{\textbf{#1}}}

\definecolor{a0}{rgb}{0.0, 0.5, 0.0}
\definecolor{bistre}{rgb}{0.24, 0.17, 0.12}
\definecolor{amethyst}{rgb}{0.6, 0.4, 0.8}
\definecolor{blue-violet}{rgb}{0.54, 0.17, 0.89}
\definecolor{Rcolor}{RGB}{150,160,190}
\definecolor{blush}{rgb}{0.87, 0.36, 0.51}
\definecolor{brightturquoise}{rgb}{0.03, 0.91, 0.87}
\definecolor{burntorange}{rgb}{0.8, 0.33, 0.0}

\begin{document}

\def\spacingset#1{\renewcommand{\baselinestretch}
{#1}\small\normalsize} \spacingset{1}

\if0\blind
{
  \title{\bf Forecasting multiple functional time series in a group structure: an application to mortality}
  \author{
    Han Lin Shang\footnote{Postal address: Research School of Finance, Actuarial Studies and Statistics, Level 4, Building 26C, Kingsley Street, Australian National University, Acton, Canberra, ACT 2601, Australia; Phone number: +61(2) 6125 0535; Fax number: +61(2) 6125 0087; Email: hanlin.shang@anu.edu.au}\\
    Research School of Finance, Actuarial Studies and Statistics \\
 Australian National University \\
\\
  Steven Haberman \\
 Cass Business School\\
  City, University of London}
\maketitle
} \fi

\if1\blind
{
  \bigskip
  \bigskip
  \bigskip
  \begin{center}
    {\LARGE\bf Forecasting multiple functional time series in a group structure: an application to mortality}
\end{center}
  \medskip
} \fi

\bigskip
\begin{abstract}
When modeling sub-national mortality rates, we should consider three features: (1) how to incorporate any possible correlation among sub-populations to potentially improve forecast accuracy through multi-population joint modeling; (2) how to reconcile sub-national mortality forecasts so that they aggregate adequately across various levels of a group structure; (3) among the forecast reconciliation methods, how to combine their forecasts to achieve improved forecast accuracy.
To address these issues, we introduce an extension of grouped univariate functional time series method.
We first consider a multivariate functional time series method to jointly forecast multiple related series. We then evaluate the impact and benefit of using forecast combinations among the forecast reconciliation methods. Using the Japanese regional age-specific mortality rates, we investigate one-step-ahead to 15-step-ahead point and interval forecast accuracies of our proposed extension and make recommendations.
\\

\noindent {Keywords}: forecast reconciliation; multivariate functional principal component analysis; bottom-up method; optimal-combination method; Japanese mortality database.
\\

\noindent JEL code: C53; C55
\end{abstract}

\noindent
\vfill

\newpage
\spacingset{1.45}

\section{Introduction}\label{sec:intro}

Increases in longevity and an ageing population have led to concerns regarding the sustainability of pensions, healthcare, and aged-care systems in many developed nations. These concerns have led to an increasing interest among government policymakers and planners to engage in the development of more accurate modeling and forecasting age-specific mortality rates. Further, annuity and pension products depend crucially on the forecast accuracy of mortality rates or their associated survival probabilities. The survival probability has been consistently underestimated in the last few decades. As a consequence, pension funds and insurance companies face longevity risk. Longevity risk is a potential risk arising from the increasing life expectancy of policyholders, which can eventually result in higher payout ratios than expected. 

Many statistical methods have been proposed for modeling and forecasting age-specific mortality rates at the national level \citep[see, e.g.,][for earlier reviews]{SBH11}. Of these, a significant milestone in demographic forecasting was the work by \cite{LC92}. They implemented a principal component analysis to model age-specific mortality and extracted a single time-varying index of the level of mortality, from which the forecasts were obtained by a random walk with drift. In the demographic literature, many extensions and modifications of the Lee-Carter method are collated in \cite{SBH11}. 

Modeling mortality at the sub-national level is not only important but also challenging.  On the one hand, the sub-national mortality rates often suffer from relatively poor data quality with possible missing data; on the other hand, sub-national forecasts of age-specific mortality rates are useful for informing regional policy and understanding the heterogeneity in the whole population. Further, improved understanding of individual characteristics will enable insurers to price more accurately annuity products for annuitants, as in the growing market for ``enhanced annuities" in the United Kingdom \citep[see, e.g.,][]{OP16}. 

To our knowledge, there are few papers that model national and sub-national mortality together with respect to a group structure. \cite{SH17} and \cite{SH17b} proposed the bottom-up and optimal combination based on ordinary least squares for reconciling forecasts in a group structure. \cite{LLL+19} applied the optimal combination based on generalized least squares of \cite{WAH19} to reconcile cause-specific mortality forecasts in a three-level hierarchy. In all these works, a demographic model, such as the Lee-Carter model or the functional time series model of \cite{HU07}, is used to forecast each series within each level of a group structure. Similar to these papers, we also consider the forecast reconciliation methods to reconcile forecasts of age-specific mortality rates, and improve the forecast accuracy. In a novel approach that differs from these papers, we jointly model and forecast sub-national age-specific mortality rates at each level of a group structure to capture correlation among the series and further improve forecast accuracy. We consider a multivariate functional time series method to forecast multiple related series jointly instead of applying a univariate functional time series method to forecast each series individually. Also, we introduce a forecast combination approach among the grouped forecasting methods to potentially improve the forecast accuracy.

Multiple-population modeling and forecasting have attracted increasing attention in actuarial science \citep[see, e.g.,][]{HH13} and demography \citep[see, e.g.,][]{LL05}. Our extension links multiple-population forecasting with grouped functional time series forecasting.
The underlying intuition is that when multiple sub-populations are correlated, the proposed multivariate functional time-series method can capture correlation among the multiple series, and in turn, can lead to improved forecast accuracy. 

The idea of forecast combination has been studied in statistics, dating back to the seminal work by \cite{BG69}. 
The underlying intuition is that when various grouped forecasting methods are combined, forecast combination may reduce bias, variance and uncertainty because of different assumptions, model structures and degrees of model complexity.

The remainder of this paper is structured as follows. In Section~\ref{sec:2}, we describe the Japanese national and sub-national mortality observed from 1975 to 2016. In Section~\ref{sec:3}, we introduce a multivariate functional time-series forecasting method for forecasting multiple series at each level of a group structure. The key technique in our multivariate functional time-series forecasting method is multivariate functional principal component analysis. In Section~\ref{sec:4}, we introduce two grouped forecasting methods and their forecast combination. In Section~\ref{sec:5}, we compare the forecast accuracy in two ways: 
\begin{inparaenum}
\item[(1)] between two reconciliation methods and their forecast combination; 
\item[(2)] between univariate and multivariate functional time-series forecasting methods. 
\end{inparaenum}
In Section~\ref{sec:6}, we introduce an actuarial application and apply the most accurate forecasting method to estimate the temporary life annuity prices for different ages and maturities. Conclusions are drawn in Section~\ref{sec:7}, along with some reflections on how the methods presented here can be further extended.

\section{Japanese age-specific mortality rates}\label{sec:2}

We study Japanese age-specific mortality rates from 1975 to 2016, obtained from the Japanese Mortality Database \citep{JMD17}. Given that our focus is on the pricing of annuities, we consider ages from 60 to 99 in a single year of age, and the last age group is the age at and beyond 100. The aggregation of the last age group is to avoid the missing data issue of those at the older ages. The structure of the data is presented in Table~\ref{tab:1}, where each row denotes a level of disaggregation.

\begin{table}[!htbp]
\tabcolsep 0.5in
\centering
\caption{Group structure of Japanese mortality rates.}\label{tab:1}
\begin{tabular}{@{}lr@{}}
\toprule
Group level & Number of series \\\midrule
Japan & 1 \\
Sex & 2 \\
Region & 8 \\
Region $\times$ Sex & 16 \\
Prefecture & 47 \\
Prefecture $\times$ Sex & 94 \\
\midrule
Total & 168 \\\bottomrule
\end{tabular}
\end{table}

At the top level, we have total age-specific mortality rates for entire Japan. We can split these total mortality rates by various attributes such as sex, region or prefecture. For this data set, there are eight regions in Japan, which contain a total of 47 prefectures. The most disaggregated data arise when we consider the mortality rates for each combination of prefecture and sex, giving a total of $47\times 2=94$ series \citep{SH17}. All in all, across all levels of disaggregation, there are 168 series. Note that the order of the disaggregation is not unique in any group structure, as we may first disaggregate series by region. The possibility of different disaggregation orders may impact forecast accuracy and we are investigating this issue in a separate project.

\section{Multivariate functional time-series forecasting}\label{sec:3}


Joint modeling mortality for two or more populations simultaneously is paramount, as it allows one to model the correlations among two or more populations, distinguish between long-term and short-term effects in the mortality evolution, and explore the additional information contained in the experience of other populations to further improve forecast accuracy. These populations can be grouped by sex, geography, ethnicity, socioeconomic status and other attributes. 

\subsection{Multivariate functional principal component analysis}

Let $y_t^{(j)}(x_i)$ be the log central mortality rates observed at the beginning of each year $t=1,2,\dots,n$ at observed ages $(x_1, x_2, \dots, x_p)$ where $x$ is a continuous age variable, $p$ denotes the number of ages, and superscript $^{(j)}$ represents $j$\textsuperscript{th} series. By applying a penalized regression spline smoothing, we obtain smoothed series, $f_t^{(j)}(x)$ that is observed at discrete data points with errors. 

As in the case of multiple subpopulations, the multivariate functional time series are combined in a vector with
\begin{equation}
\bm{f}(x) = \left[f^{(1)}(x),\dots,f^{(\omega)}(x)\right]\in R^{\omega}.
\end{equation}
These multivariate functions are defined over the same domain $\Gamma$. The common domain $\Gamma$ must be compact sets in $R^{\omega}$, $\omega\in N$ with finite measure and with each element $f^{(j)}(x)$ assumed to be a square-integrable function $\mathcal{L}^2(\Gamma)$, where $j=1,\dots,\omega$. For mathematical convenience, we let
\begin{equation}
\bm{\mu}(x) :=\mathbb{E}\left[\bm{f}(x)\right] = \left\{\mathbb{E}\left[f^{(1)}(x)\right], \dots, \mathbb{E}\left[f^{(\omega)}(x)\right]\right\} = \bm{0},
\end{equation}
where $\bm{0}$ denotes a vector of zeros. For $x, z\in \Gamma$, the cross-covariance function is defined with elements
\begin{equation}
\mathcal{K}_{lj}(x, z) := \mathbb{E}\left[f^{(l)}(x)f^{(j)}(z)\right] = \text{Cov}\left[f^{(l)}(x), f^{(j)}(z)\right]. \label{eq:cov}
\end{equation}
From the cross-covariance function, we can deduce the eigenfunction as
\begin{equation}
\left(\mathcal{K}\phi\right)^{(l)}(z) = \sum^{\omega}_{j=1}\int_{\Gamma}\mathcal{K}_{lj}(x, z)\phi^{(j)}(x)dx,
\end{equation}
where $\mathcal{K}$ induces the kernel of an integral operator, $\phi \mapsto \mathcal{K}\phi$ is a square-integrable function.

From the cross-covariance function, there exists an orthonormal sequence $(\phi_k)$ of continuous functions in $\mathcal{L}^2(\Gamma)$ and a non-increasing sequence, $\lambda_k$, of positive numbers, such that
\begin{equation}
\mathcal{K}_{lj}(x, z) = \sum^{\infty}_{k=1}\lambda_k \phi_k^{(l)}(x)\phi_k^{(j)}(z), \qquad x, z\in \mathcal{I}.
\end{equation}

By functional principal component analysis, a de-centered stochastic process $f_t^{(j)}(x)$ can be expressed as
\begin{equation}
f_t^{(j)}(x) = \sum^{\infty}_{k=1}\beta^{(j)}_{t,k}\phi_k^{(j)}(x) \approx  \sum_{k=1}^{K}\beta^{(j)}_{t,k}\phi^{(j)}_k(x), \label{eq:FPCA}
\end{equation}
where $\big\{\phi_1^{(j)}(x),\dots,\phi_{K}^{(j)}(x)\big\}$ is a set of the first $K$ functional principal components for the $j$\textsuperscript{th} subpopulation; $\bm{\beta}_1^{(j)} = \big(\beta_{1,1}^{(j)},\dots,\beta_{n,1}^{(j)}\big)^{\top}$ and $\big\{\bm{\beta}^{(j)}_1,\dots,\bm{\beta}^{(j)}_{K}\big\}$ denotes a set of principal component scores and $\bm{\beta}_k^{(j)}\sim N\big(0,\lambda_k^{(j)}\big)$ where $\lambda_k^{(j)}$ is the $k^{\text{th}}$ eigenvalue of the covariance function for the $j^{\text{th}}$ subpopulation in~\eqref{eq:cov}; and $K<n$ is the retained number of functional principal components. Expansion~\eqref{eq:FPCA} facilitates dimension reduction as the first $K$ terms often provide a good approximation to the infinite sum, and thus the information contained in $\bm{f}^{(j)}(x)=[f^{(j)}_1(x),\dots,f^{(j)}_n(x)]$ can be adequately summarized by the $K$-dimensional vector $\big(\bm{\beta}^{(j)}_1,\dots,\bm{\beta}^{(j)}_{K}\big)$. 

The optimal value of $K$ can be selected by a ratio method:
\begin{equation}
\argmin_{1\leq K\leq n-1}\left|\frac{\widehat{\lambda}_{K+1}}{\widehat{\lambda}_{K}}\right| \label{eq:ratio}
\end{equation}
or the optimal value of $K$ can be selected by explaining at least 90\% of total variation:
\begin{equation}
\argmin_{1\leq K\leq n}\left\{\frac{\sum^K_{k=1}\widehat{\lambda}_K}{\sum^n_{k=1}\widehat{\lambda}_K}\geq 0.9\right\}. \label{eq:CPV}
\end{equation}
We take the maximum of the $K$ values obtained from~\eqref{eq:ratio} and~\eqref{eq:CPV}.

The matrix formulation of~\eqref{eq:FPCA} is
\begin{equation}
\bm{f}_t(x) \approx \bm{\beta}_t\bm{\Phi}^{\top},
\end{equation}
where $\bm{f}_t(x) = \left[f_t^{(1)}(x), f_t^{(2)}(x), \dots,f_t^{(\omega)}(x)\right]$, $\bm{\beta}_t = \left[\beta_{t,1}^{(1)}, \dots, \beta_{t,K}^{(1)}, \beta_{t,1}^{(2)},\dots,\beta_{t,K}^{(2)},\dots, \beta_{t,1}^{(\omega)},\dots,\beta_{t,K}^{(\omega)}\right]$ being the vector of the basis expansion coefficients, and 
\begin{equation}
\bm{\Phi}(x) = \left( \begin{array}{ccccccccc}
\phi_1^{(1)}(x) & \cdots & \phi_{K}^{(1)}(x) & 0 & \cdots & 0 & 0 & \cdots & 0 \\
0 & \cdots & 0 & \phi_1^{(2)}(x) & \cdots & \phi_{K}^{(2)}(x) & 0 & \cdots & 0 \\
\vdots & \vdots & \vdots & \vdots & \vdots & \vdots & \vdots & \vdots & \vdots \\
 0 & \cdots & 0 & 0 & \cdots & 0 & \phi_1^{(\omega)}(x) & \cdots & \phi_{K}^{(\omega)}(x)  \end{array} \right)_{\omega\times \ell},
\end{equation}
where $\ell = \omega \times K$.

The advantage of our multivariate functional time-series forecasting method is that the correlations among sub-populations can be captured in the cross-covariance structure described in~\eqref{eq:cov}. The disadvantage of our proposal is that we implicitly assume that all series share the same retained number of functional principal components, and consequently we may lose forecast accuracy for a particular series. 

Based on the estimated covariance function, we can extract empirical functional principal component functions $\bm{\mathcal{B}}=\{\widehat{\phi}^{(j)}_1(x), \dots, \widehat{\phi}^{(j)}_{K}(x)\}$ using functional principal component analysis. Conditioning on the smoothed functions $\bm{f}^{(j)}(x) = \{f_1^{(j)}(x),\dots, f_n^{(j)}(x)\}$ and the estimated functional principal components $\bm{\mathcal{B}}$, the $h$-step-ahead point forecast of $f_{n+h}^{(j)}(x)$ can be obtained as
\begin{equation}
\widehat{y}_{n+h|n}^{(j)}(x) = \widehat{f}_{n+h|n}^{(j)}(x) = \mathbb{E}\left[f_{n+h}^{(j)}(x)\big|\bm{f}^{(j)}(x), \bm{\mathcal{B}}\right] = \sum_{k=1}^{K}\widehat{\beta}^{(j)}_{n+h|n,k}\widehat{\phi}^{(j)}_k(x),
\end{equation}
where $\widehat{\beta}_{n+h|n, k}^{(j)}$ represents the time-series forecasts of the $k$\textsuperscript{th} principal component scores for the $j$\textsuperscript{th} subpopulation, which can be obtained by using autoregressive integrated moving average models.

\section{Grouped forecasting methods}\label{sec:4}

\subsection{Notation}

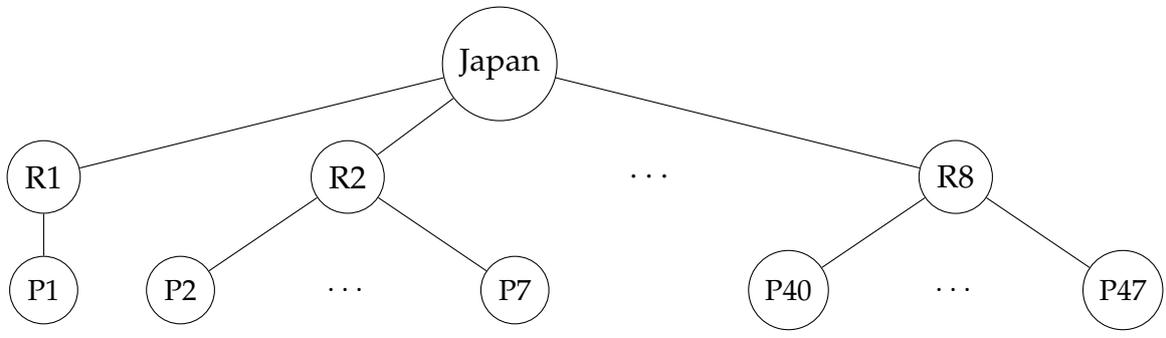
\begin{figure}[!htb]
\centering\begin{tikzpicture}
\tikzstyle{every node}=[minimum size = 8mm]
\tikzstyle[level distance=10cm] \tikzstyle[sibling distance=40cm]
\tikzstyle{level 3}=[sibling distance=16mm,font=\footnotesize]
\tikzstyle{level 2}=[sibling distance=22mm,font=\small]
\tikzstyle{level 1}=[sibling distance=40mm,font=\normalsize]
\node[circle,draw]{Japan}
   child {node[circle,draw] {R1}
   	     child {node[circle,draw] {P1}}}
   child {node[circle,draw] {R2}
   		child {node[circle,draw] {P2}}
      child {node {$\cdots$}edge from parent[draw=none]}
		child {node[circle,draw] {P7}}
		}
  child {node {$\cdots$}edge from parent[draw=none]}
   child {node[circle,draw] {R8}
   		child{node[circle,draw] {P40}}
	     child{node {$\cdots$}edge from parent[draw=none]}
  	child{node[circle,draw] {P47}}
 };
\end{tikzpicture}
\medskip
\caption{The Japanese geographical group structure tree diagram, with eight regions and 47 prefectures -- each node has the female, male and total age-specific mortality rates.}\label{fig:2}
\end{figure}

We introduce the grouped forecasting methods using the Japanese age-specific mortality rates provided in Section~\ref{sec:2}. The Japanese data follow a three-level geographic group structure, coupled with a sex-grouping variable \citep{SH17}. The geographical group structure is presented in Figure~\ref{fig:2}. Japan can be split into eight regions from north to south, which is then divided into 47 prefectures \citep{SH17}.

The data can also be split by sex. Each of the nodes in the geographical group structure can also be split into males and females. We refer to a particular disaggregated series using the notation $X\ast S$, referring to the geographic area $X$ and the sex $S$, where $X$ can take the values shown in Figure~\ref{fig:2} and $S$ can take values M (males), F (females) or T (total). For example, R1$\ast$ F denotes females in Region 1; P1$\ast$ T denotes all females and males in Prefecture 1; Japan $\ast$ M denotes all males in Japan.

Denote E$_{X\ast S, t}(x)$ as the exposure-to-risk for series $X\ast S$ in year $t$ and age $x$, and let $\text{D}_{X\ast S, t}(x)$ be the number of deaths for series $X\ast S$ in year $t$ and age $x$. The age-specific mortality rate is $\text{R}_{X\ast S, t}(x) = \text{D}_{X\ast S, t}(x)/\text{E}_{X\ast S, t}(x)$. 

To simplify expressions, we will drop the age argument $(x)$. Then for a given age, we can write
\[
\hspace{-.2in} \underbrace{ \left[
\begin{footnotesize}
\begin{array}{l}
\text{R}_{\text{Japan}\ast\text{T},t} \\
\text{R}_{\textcolor{red}{\text{Japan}\ast\text{F},t}} \\
\text{R}_{\textcolor{red}{\text{Japan}\ast\text{M},t}} \\
\text{R}_{\textcolor{a0}{\text{R1}\ast\text{T},t}} \\
\text{R}_{\textcolor{a0}{\text{R2}\ast\text{T},t}} \\
\vdots \\
\text{R}_{\textcolor{a0}{\text{R8}\ast\text{T},t}} \\
\text{R}_{\textcolor{blue-violet}{\text{R1}\ast\text{F},t}} \\
\text{R}_{\textcolor{blue-violet}{\text{R2}\ast\text{F},t}} \\
\vdots \\
\text{R}_{\textcolor{blue-violet}{\text{R8}\ast\text{F},t}} \\
\text{R}_{\textcolor{burntorange}{\text{R1}\ast\text{M},t}} \\
\text{R}_{\textcolor{burntorange}{\text{R2}\ast\text{M},t}} \\
\vdots \\
\text{R}_{\textcolor{burntorange}{\text{R8}\ast\text{M},t}} \\
\text{R}_{\textcolor{blue}{\text{P1}\ast\text{T},t}} \\
\text{R}_{\textcolor{blue}{\text{P2}\ast\text{T},t}} \\
\vdots \\
\text{R}_{\textcolor{blue}{\text{P47}\ast\text{T},t}} \\
\text{R}_{\textcolor{purple}{\text{P1}\ast\text{F},t}} \\
\text{R}_{\textcolor{purple}{\text{P1}\ast\text{M},t}} \\
\text{R}_{\textcolor{purple}{\text{P2}\ast\text{F},t}} \\
\text{R}_{\textcolor{purple}{\text{P2}\ast\text{M},t}} \\
\vdots \\
\text{R}_{\textcolor{purple}{\text{P47}\ast\text{F},t}} \\
\text{R}_{\textcolor{purple}{\text{P47}\ast\text{M},t}} \\ \end{array}
\end{footnotesize} \right]}_{\bm{R}_t} =
\underbrace{\left[
\begin{footnotesize}
\begin{array}{ccccccccccc}
\frac{\text{E}_{\text{P1}\ast\text{F},t}}{\text{E}_{\text{Japan}\ast \text{T},t}} & \frac{\text{E}_{\text{P1}\ast \text{M},t}}{\text{E}_{\text{Japan}\ast \text{T},t}} & \frac{\text{E}_{\text{P2}\ast\text{F},t}}{\text{E}_{\text{Japan}\ast\text{T},t}} & \frac{\text{E}_{\text{P2}\ast\text{M},t}}{\text{E}_{\text{Japan}\ast\text{T},t}}  & \frac{\text{E}_{\text{P3}\ast\text{F},t}}{\text{E}_{\text{Japan}\ast\text{T},t}} & \frac{\text{E}_{\text{P3}\ast\text{M},t}}{\text{E}_{\text{Japan}\ast\text{T},t}} & \cdots & \frac{\text{E}_{\text{P47}\ast\text{F},t}}{\text{E}_{\text{Japan}\ast\text{T},t}} & \frac{\text{E}_{\text{P47}\ast\text{M},t}}{\text{E}_{\text{Japan}\ast\text{T},t}} \\
\textcolor{red}{\frac{\text{E}_{\text{P1}\ast\text{F},t}}{\text{E}_{\text{Japan}\ast\text{F},t}}} & \textcolor{red}{0} & \textcolor{red}{\frac{\text{E}_{\text{P2}\ast\text{F},t}}{\text{E}_{\text{Japan}\ast\text{F},t}}} & \textcolor{red}{0} & \textcolor{red}{\frac{\text{E}_{\text{P3}\ast\text{F},t}}{\text{E}_{\text{Japan}\ast\text{F},t}}} & \textcolor{red}{0} & \cdots & \textcolor{red}{\frac{\text{E}_{\text{P47}\ast\text{F},t}}{\text{E}_{\text{Japan}\ast\text{F},t}}} & \textcolor{red}{0} \\
\textcolor{red}{0} & \textcolor{red}{\frac{\text{E}_{\text{P1}\ast\text{M},t}}{\text{E}_{\text{Japan}\ast\text{M},t}}}  & \textcolor{red}{0} & \textcolor{red}{\frac{\text{E}_{\text{P2}\ast\text{M},t}}{\text{E}_{\text{Japan}\ast\text{M},t}}} & \textcolor{red}{0} & \textcolor{red}{\frac{\text{E}_{\text{P3}\ast\text{M},t}}{\text{E}_{\text{Japan}\ast\text{M},t}}} & \cdots & \textcolor{red}{0} & \textcolor{red}{\frac{\text{E}_{\text{P47}\ast\text{M},t}}{\text{E}_{\text{Japan}\ast\text{M},t}}} \\
\textcolor{a0}{\frac{\text{E}_{\text{P1}\ast\text{F},t}}{\text{E}_{\text{R1}\ast\text{T},t}}} & \textcolor{a0}{\frac{\text{E}_{\text{P1}\ast\text{M},t}}{\text{E}_{\text{R1}\ast\text{T},t}}} & \textcolor{a0}{0} & \textcolor{a0}{0} & \textcolor{a0}{0} & \textcolor{a0}{0} & \cdots  & \textcolor{a0}{0} & \textcolor{a0}{0} \\
\textcolor{a0}{0} & \textcolor{a0}{0} & \textcolor{a0}{\frac{\text{E}_{\text{P2}\ast\text{F},t}}{\text{E}_{\text{R2}\ast\text{T},t}}} & \textcolor{a0}{\frac{\text{E}_{\text{P2}\ast\text{M},t}}{\text{E}_{\text{R2}\ast\text{T},t}}} & \textcolor{a0}{\frac{\text{E}_{\text{P3}\ast\text{F},t}}{\text{E}_{\text{R2}\ast\text{T},t}}} & \textcolor{a0}{\frac{\text{E}_{\text{P3}\ast\text{M},t}}{\text{E}_{\text{R2}\ast\text{T},t}}} & \cdots & \textcolor{a0}{0} & \textcolor{a0}{0} \\
\vdots & \vdots & \vdots & \vdots & \vdots & \vdots & \cdots & \vdots & \vdots \\
\textcolor{a0}{0} & \textcolor{a0}{0} & \textcolor{a0}{0} & \textcolor{a0}{0} & \textcolor{a0}{0} & \textcolor{a0}{0} & \cdots & \textcolor{a0}{\frac{\text{E}_{\text{P47}\ast\text{F},t}}{\text{E}_{\text{R8}\ast\text{T},t}}} & \textcolor{a0}{\frac{\text{E}_{\text{P47}\ast\text{M},t}}{\text{E}_{\text{R8}\ast\text{T},t}}} \\
\textcolor{blue-violet}{\frac{\text{E}_{\text{P1}\ast\text{F},t}}{\text{E}_{\text{R1}\ast\text{F},t}}} & \textcolor{blue-violet}{0} & \textcolor{blue-violet}{0} & \textcolor{blue-violet}{0} & \textcolor{blue-violet}{0} & \textcolor{blue-violet}{0} &  \cdots & \textcolor{blue-violet}{0} & \textcolor{blue-violet}{0} \\
\textcolor{blue-violet}{0} & \textcolor{blue-violet}{0} & \textcolor{blue-violet}{\frac{\text{E}_{\text{P2}\ast\text{F},t}}{\text{E}_{\text{R2}\ast\text{F},t}}} & \textcolor{blue-violet}{0} & \textcolor{blue-violet}{\frac{\text{E}_{\text{P3}\ast\text{F},t}}{\text{E}_{\text{R2}\ast\text{F},t}}} & \textcolor{blue-violet}{0} & \cdots & \textcolor{blue-violet}{0} & \textcolor{blue-violet}{0}  \\
\vdots & \vdots & \vdots & \vdots & \vdots & \vdots & \cdots & \vdots & \vdots \\
\textcolor{blue-violet}{0} & \textcolor{blue-violet}{0}  & \textcolor{blue-violet}{0}  & \textcolor{blue-violet}{0}  & \textcolor{blue-violet}{0}  & \textcolor{blue-violet}{0}  & \cdots & \textcolor{blue-violet}{\frac{\text{E}_{\text{P47}\ast\text{F},t}}{\text{E}_{\text{R8}\ast\text{F},t}}} & \textcolor{blue-violet}{0}\\
\textcolor{burntorange}{0} & \textcolor{burntorange}{\frac{\text{E}_{\text{P1}\ast\text{M},t}}{\text{E}_{\text{R1}\ast\text{M},t}}} & \textcolor{burntorange}{0} &\textcolor{burntorange}{0} & \textcolor{burntorange}{0} & \textcolor{burntorange}{0} & \cdots & \textcolor{burntorange}{0} & \textcolor{burntorange}{0} \\
\textcolor{burntorange}{0} & \textcolor{burntorange}{0} & \textcolor{burntorange}{0} & \textcolor{burntorange}{\frac{\text{E}_{\text{P2}\ast\text{M},t}}{\text{E}_{\text{R2}\ast\text{M},t}}} & \textcolor{burntorange}{0} & \textcolor{burntorange}{\frac{\text{E}_{\text{P3}\ast\text{M},t}}{\text{E}_{\text{R2}\ast\text{M},t}}} & \cdots & \textcolor{burntorange}{0} & \textcolor{burntorange}{0} \\
\vdots & \vdots & \vdots & \vdots & \vdots & \vdots & \cdots & \vdots & \vdots \\
\textcolor{burntorange}{0} & \textcolor{burntorange}{0} & \textcolor{burntorange}{0} & \textcolor{burntorange}{0} & \textcolor{burntorange}{0} & \textcolor{burntorange}{0} & \cdots & \textcolor{burntorange}{0} & \textcolor{burntorange}{\frac{\text{E}_{\text{P47}\ast\text{M},t}}{\text{E}_{\text{R8}\ast\text{M},t}}} \\
\textcolor{blue}{\frac{\text{E}_{\text{P1}\ast\text{F},t}}{\text{E}_{\text{P1}\ast\text{T},t}}} & \textcolor{blue}{\frac{\text{E}_{\text{P1}\ast\text{M},t}}{\text{E}_{\text{P1}\ast\text{T},t}}} & \textcolor{blue}{0} & \textcolor{blue}{0} & \textcolor{blue}{0} & \textcolor{blue}{0} & \cdots & \textcolor{blue}{0} & \textcolor{blue}{0} \\
\textcolor{blue}{0} & \textcolor{blue}{0}  &  \textcolor{blue}{\frac{\text{E}_{\text{P2}\ast\text{F},t}}{\text{E}_{\text{P2}\ast\text{T},t}}} & \textcolor{blue}{\frac{\text{E}_{\text{P2}\ast\text{M},t}}{\text{E}_{\text{P2}\ast\text{T},t}}} & \textcolor{blue}{0} & \textcolor{blue}{0}  & \cdots & \textcolor{blue}{0} & \textcolor{blue}{0} \\
\vdots & \vdots & \vdots & \vdots & \vdots & \vdots & \cdots & \vdots & \vdots \\
\textcolor{blue}{0} & \textcolor{blue}{0} & \textcolor{blue}{0} & \textcolor{blue}{0} & \textcolor{blue}{0} & \textcolor{blue}{0} & \cdots & \textcolor{blue}{\frac{\text{E}_{\text{P47}\ast\text{F},t}}{\text{E}_{\text{P47}\ast\text{T},t}}} & \textcolor{blue}{\frac{\text{E}_{\text{P47}\ast\text{M},t}}{\text{E}_{\text{P47}\ast\text{T},t}}} \\
\textcolor{purple}{1} & \textcolor{purple}{0} & \textcolor{purple}{0} & \textcolor{purple}{0} & \textcolor{purple}{0} & \textcolor{purple}{0} & \cdots & \textcolor{purple}{0} & \textcolor{purple}{0} \\
\textcolor{purple}{0} & \textcolor{purple}{1} & \textcolor{purple}{0} & \textcolor{purple}{0} & \textcolor{purple}{0} & \textcolor{purple}{0} & \cdots & \textcolor{purple}{0} & \textcolor{purple}{0} \\
\textcolor{purple}{0} & \textcolor{purple}{0} & \textcolor{purple}{1} & \textcolor{purple}{0} & \textcolor{purple}{0} & \textcolor{purple}{0} & \cdots & \textcolor{purple}{0} & \textcolor{purple}{0} \\
\textcolor{purple}{0} & \textcolor{purple}{0} & \textcolor{purple}{0} & \textcolor{purple}{1} & \textcolor{purple}{0} & \textcolor{purple}{0} & \cdots & \textcolor{purple}{0} & \textcolor{purple}{0} \\
\vdots & \vdots & \vdots & \vdots & \vdots & \vdots & \cdots & \vdots & \vdots  \\
\textcolor{purple}{0} & \textcolor{purple}{0} & \textcolor{purple}{0} & \textcolor{purple}{0} & \textcolor{purple}{0} & \textcolor{purple}{0} & \cdots  & \textcolor{purple}{1} & \textcolor{purple}{0}\\
\textcolor{purple}{0} & \textcolor{purple}{0} & \textcolor{purple}{0} & \textcolor{purple}{0} & \textcolor{purple}{0} & \textcolor{purple}{0} & \cdots & \textcolor{purple}{0} & \textcolor{purple}{1} \\
\end{array}
\end{footnotesize} \right]}_{\bm{S}_t}
\underbrace{\left[
\begin{footnotesize}
\begin{array}{l}
\text{R}_{\text{P1}\ast\text{F},t} \\
\text{R}_{\text{P1}\ast\text{M},t} \\
\text{R}_{\text{P2}\ast\text{F},t} \\
\text{R}_{\text{P2}\ast\text{M},t} \\
\vdots \\
\text{R}_{\text{P47}\ast\text{F},t} \\
\text{R}_{\text{P47}\ast\text{M},t} \\ 
\end{array}
\end{footnotesize}
\right]}_{\bm{b}_t}
\]
or $\bm{R}_t = \bm{S}_t\bm{b}_t$, where $\bm{R}_t$ is a vector containing all series at all levels of disaggregation, $\bm{b}_t$ is a vector of the most disaggregated series, and $\bm{S}_t$ shows how the two are connected. 


\subsection{Bottom-up method}\label{sec:BU}

As the simplest grouped forecasting method, the bottom-up method first generates independent forecasts for each series at the most disaggregated level, and then aggregates these to produce all of the required forecasts \citep{SH17}. For example, reverting to the Japanese data, we first generate $h$-step-ahead independent forecasts for the most disaggregated series, namely $\widehat{b}_{n+h}=\big[\widehat{R}_{\text{P1}\ast \text{F}, n+h}, \widehat{R}_{\text{P1}\ast \text{M}, n+h}, \dots,$ $\widehat{R}_{\text{P47}\ast \text{F}, n+h}, \widehat{R}_{\text{P47}\ast \text{M}, n+h}\big]^{\top}$. Then, we obtain forecasts for all series as
\begin{equation}
\overline{\bm{R}}_{n+h}^{\text{BU}} = \bm{S}_{n+h}\widehat{\bm{b}}_{n+h},
\end{equation}
where $\overline{\bm{R}}_{n+h}^{\text{BU}}$ denotes the reconciled forecasts obtained from the bottom-up method. 

The bottom-up method performs well when the bottom-level series have a high signal-to-noise ratio. In contrast, the bottom-up method may lead to inaccurate forecasts of the top-level series, in particular when there are missing or noisy data at the bottom level \citep[see, e.g.,][]{SH17, SH17b}.

\subsection{Optimal-combination method}\label{sec:OC}

Instead of considering only the bottom-level series, \cite{HAA+11} have proposed the optimal-combination method where independent forecasts for all series are computed independently, and then the resultant forecasts are reconciled so that they satisfy the aggregation constraints via the summing matrix. The optimal-combination method combines the independent forecasts through linear regression by generating a set of revised forecasts that are as close as possible to the independent forecasts, but that also aggregate consistently within the group. The method is derived by expressing the independent forecasts as the response variable of the linear regression
\begin{equation}
\widehat{\bm{R}}_{n+h} = \bm{S}_{n+h}\bm{\beta}_{n+h}+\epsilon_{n+h},
\end{equation}
where $\widehat{\bm{R}}_{n+h}$ is a matrix of $h$-step-ahead independent forecasts for all series, stacked in the same order as for the original data; $\bm{\beta}_{n+h} = \mathbb{E}[\bm{b}_{n+h}|\bm{R}_1,\dots,\bm{R}_n]$ is the unknown mean of the independent forecasts of the most disaggregated series; and $\bm{\epsilon}_{n+h}$ represents the reconciliation errors.

To estimate the regression coefficient, \cite{HAA+11} have proposed an ordinary least-squares solution,
\begin{equation}
\widehat{\bm{\beta}}_{n+h} = \left(\bm{S}_{n+h}^{\top}\bm{S}_{n+h}\right)^{-1}\bm{S}_{n+h}^{\top}\widehat{\bm{R}}_{n+h}.
\end{equation}
The revised forecasts are given by
\begin{equation}
\overline{\bm{R}}_{n+h}^{\text{OLS}} = \bm{S}_{n+h}\widehat{\bm{\beta}}_{n+h} = \bm{S}_{n+h}\left(\bm{S}_{n+h}^{\top}\bm{S}_{n+h}\right)^{-1}\bm{S}_{n+h}^{\top}\widehat{\bm{R}}_{n+h}.
\end{equation}

\subsection{Forecast combination}\label{sec:4.4}

Forecast combination involves the computation of weighted means for the forecast. The averaged forecast for horizon $h$ is computed as the weighted mean:
\begin{equation}
\overline{\bm{R}}_{n+h}^{\text{comb}} =\sum^G_{g=1}w_g \overline{\bm{R}}_{n+h}^{g},
\end{equation}
where $\overline{\bm{R}}_{n+h}^g$ denotes forecasts obtained from a grouped forecasting method, such as the bottom-up or optimal-combination method, and $\overline{\bm{R}}_{n+h}^{\text{comb}}$ represents the model-averaged forecast; and $(w_1, w_2,\dots, w_G)$ are weights that sum to 1. 

The crux of the problem lies in the selection of weights.  Recent studies found that past performance information enables the use of unequal weighting combinations, but simple combination methods have been shown to be robust in many settings. Here, we consider two heuristics for combining the midpoint (i.e., point forecast) and endpoints (i.e., interval forecasts) \citep[see][for other heuristics]{GTW17}. First, we combine point and interval forecasts obtained from two grouped forecasting methods with equal weighting. Second, we consider an envelope of prediction intervals and simple averaging of midpoints. 

\begin{description}
\item Average (Av). From the viewpoint of combining interval forecasts, $\overline{\bm{R}}_{n+h, L}^{\text{comb, Av}}=\frac{1}{G}\sum^G_{g=1}\overline{\bm{R}}_{n+h, L}^g$ and $\overline{\bm{R}}_{n+h, U}^{\text{comb, Av}} = \frac{1}{G}\sum^G_{g=1}\overline{\bm{R}}_{n+h, U}^g$, where $[\overline{\bm{R}}_{n+h, L}^g, \overline{\bm{R}}_{n+h, U}^g]$ for $g=1,\dots,G$ denote the lower and upper bounds for a random variable $\overline{\bm{R}}_{n+h}$, and $G$ denotes the number of grouped forecasting methods. This heuristic takes a simple average of the midpoint and endpoints. For combining point and interval forecasts, simple averages are often considered as the benchmark because of their simplicity, good performance, and robustness. When the lower and upper bounds are symmetric, Av combination method corresponds to averaging quantiles, which bridges the gap between interval forecast combination and quantile averaging \citep[see, e.g,][]{GTW17}. 
\item Envelope (En) of endpoints and simple averaging of midpoints. The individual method tends to be overconfident, so an envelope method combines the prediction intervals by taking an extreme viewpoint. The combined $\big(\overline{\bm{R}}_{n+h, L}^{\text{comb, En}}, \overline{\bm{R}}_{n+h, U}^{\text{comb, En}}\big)$ are obtained by $\overline{\bm{R}}_{n+h, L}^{\text{comb, En}}=\min(\overline{\bm{R}}_{n+h, L}^{1}, \dots, \overline{\bm{R}}_{n+h, L}^{G})$ and $\overline{\bm{R}}_{n+h, U}^{\text{comb, En}}=\max(\overline{\bm{R}}_{n+h, U}^1,\dots,\overline{\bm{R}}_{n+h, U}^G)$. This heuristic is conservative but can overcome the overconfident issue, as each method only presents a partial view of reality. The simple average of the lower and upper bounds obtained from $G$ number of methods provides a combined point forecast, where 
\begin{equation}
\overline{\bm{R}}_{n+h}^{\text{comb, AvInt}} = \frac{1}{G}\sum^G_{g=1}\frac{\overline{\bm{R}}_{n+h, L}^g + \overline{\bm{R}}_{n+h, U}^g}{2}.
\end{equation}
\end{description}

\subsection{Forecast exposure-to-risk}

Since the bottom-level forecasts are mortality rates, we ought to take into account forecast exposure-to-risk in order to reconcile death counts with respect to the group structure. The observed ratios that form the $S_t$ summing matrix are forecast using the automatic ARIMA algorithm of \cite{HK08}, when age $x=60$. For age above 60, we assume the exposure-to-risk of age $x+1$ in year $t+1$ will be the same as the exposure-to-risk of age $x$ in year $t$ \citep[see also][]{SH17}. For example, let $p_t = \text{E}_{\text{P1}\ast \text{F}, t}/\text{E}_{\text{Japan}\ast \text{T}, t}$ be a non-zero element of $S_t$. Given that we have observed $\{p_1,\dots, p_n\}$, an $h$-step-ahead forecast $\widehat{p}_{n+h}$ can be obtained. The forecasts of ratios between any two exposure-to-risks are used to form the matrix $\bm{S}_{n+h}$. To ensure summability to 1 in each row of the group structure, every non-zero ratio is normalized by dividing the sum of ratios in each row.

The potential improvement in forecast accuracy of the reconciliation methods partially relies on the accurate forecast of the $S$ matrix. Recall that the $S$ matrix includes ratios of forecast exposure-at-risk. Our cohort assumption is reasonable because it allows us to forecast ratios and populate the $S$ matrix. \cite{SH17} and \cite{SH17b} compare point forecast and interval forecast accuracies between the reconciliation methods, with the forecast $S$ matrix and actual holdout $S$ matrix, and found that it is advantageous to use the forecast $S$ matrix.

\section{Results}\label{sec:5}

\subsection{Multivariate functional time-series model fitting}\label{sec:5.0}

For the national and sub-national mortality rates, we examine the goodness-of-fit of the proposed multivariate functional time-series method to the observed data. Because the mortality rates for subpopulations may have different mean and variance terms, we standardized age-specific mortality rates by subtracting the mean function and dividing the standard deviation function before implementing a functional principal component analysis to a stacked data matrix of size $n\times (p\times \omega)$. 
For the Hokkaido data, the selected number of components is one (this is effectively a Lee-Carter type model). 

\begin{figure}[!htbp]
\centering
\includegraphics[width=\textwidth]{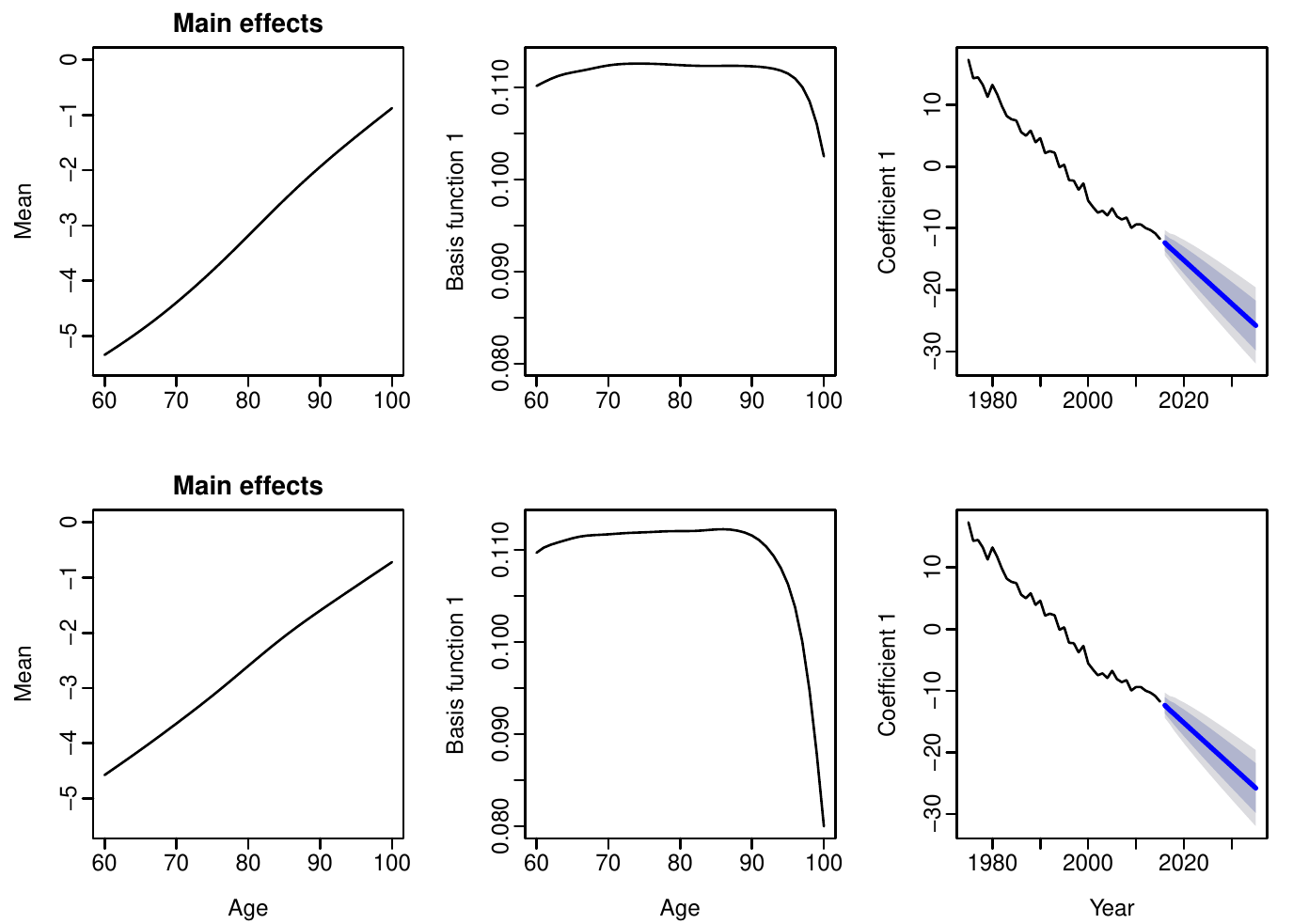}
\caption{In the top and bottom panels, multivariate functional principal component decomposition for the female and male smoothed log mortality rates in Hokkaido, respectively. In the third column, the solid blue line represents the point forecasts of principal component scores, where the dark and light grey regions represent the 80\% and 95\% pointwise prediction intervals, respectively. Standardization was applied to each subpopulation, prior to implementing the multivariate functional principal component analysis.}\label{fig:1}
\end{figure}

In the first column of Figure~\ref{fig:1}, we present the mean functions of the female and male smoothed log mortality rates. In the second and third columns, we present the first functional principal component, which accounts for around 88\% of the total variation in the joint female and male sub-populations. Given that the principal component scores are surrogates for the original functional time series, they are forecast to continue to decrease over the next 20 years. We note that the first functional principal component models the female and male mortality data at younger and older ages. 

We measure goodness-of-fit via a functional version of the $R^2$ criterion. It is given as
\begin{equation}
R^2 = 1 - \frac{\sum^n_{t=1}\int_{x\in \mathcal{I}}\left[\exp^{y_t^{(j)}(x)} - \exp^{\widehat{f}_t^{(j)}(x)}\right]^2dx}{\sum^n_{t=1}\int_{x\in \mathcal{I}}\left[\exp^{y_t^{(j)}(x)} - \exp^{\overline{y}^{(j)}(x)}\right]^2dx},\label{eq:R2}
\end{equation}
where $y_t^{(j)}(x)$ denotes the observed age-specific log mortality rates at each $t$ for the $j$\textsuperscript{th} subpopulation, $\widehat{f}_t^{(j)}(x)$ denotes the fitted age-specific log mortality rates. The larger the $R^2$ value is, the better is the goodness-of-fit by a model. It is possible for the $R^2$ criterion to take negative values. A negative $R^2$ value implies that the fitted model may not well explain the raw data that are likely to contain a large amount of measurement errors. From a negative $R^2$ value, we can quantify the amount of measurement errors exhibited in a data set and the degree of smoothing that the raw mortality data require. 

In Table~\ref{tab:2}, we report the goodness-of-fit of the univariate and multivariate functional time-series methods, as measured by the $R^2$ criterion defined in~\eqref{eq:R2}. For both female and male series, the multivariate functional time-series method generally achieves a better goodness-of-fit result than the univariate functional time-series method. The superiority of the multivariate functional time series method is because it can incorporate correlation between multiple series. This correlation provides additional information that is not possessed by the univariate functional time series method.

\begin{onehalfspace}
\begin{center}
\tabcolsep 0.05in
\begin{longtable}{@{}lrrrrlrrrr@{}}
\caption{Goodness-of-fit as measured by the $R^2$ criterion for each national and sub-national female and male age-specific mortality rates in Japan. Let FTS denote functional time series. For each series and sex, we highlight in bold the method with a higher $R^2$ value.}\label{tab:2}\\
\toprule
& \multicolumn{2}{c}{Univariate FTS} & \multicolumn{2}{c}{Multivariate FTS} & & \multicolumn{2}{c}{Univariate FTS} & \multicolumn{2}{c}{Multivariate FTS} \\ 
Series  & Female & Male & Female & Male &  Series & Female & Male & Female & Male \\\midrule
\endfirsthead
\toprule
& \multicolumn{2}{c}{Univariate FTS} & \multicolumn{2}{c}{Multivariate FTS} & & \multicolumn{2}{c}{Univariate FTS} & \multicolumn{2}{c}{Multivariate FTS} \\ 
Series  & Female & Male & Female & Male &  Series & Female & Male & Female & Male \\\midrule 
\endhead
\hline \multicolumn{10}{r}{{Continued on next page}}
\endfoot
\endlastfoot
 Japan & 0.939 & 0.802 & \textBF{0.964} & \textBF{0.902} & Mie & 0.399 & 0.110 & \textBF{0.453} & \textBF{0.208} \\ 
  Hokkaido & 0.624 & 0.262 & \textBF{0.625} & \textBF{0.369} & Shiga & 0.250 & -0.044 & \textBF{0.330} & \textBF{-0.023} \\  
  Aomori & 0.172 & 0.006 & \textBF{0.202} & \textBF{0.040} & Kyoto & 0.518 & 0.121 & \textBF{0.577} & \textBF{0.162} \\ 
  Iwate & 0.309 & 0.082 & \textBF{0.364} & \textBF{0.160} & Osaka & 0.692 & 0.246 & \textBF{0.740} & \textBF{0.288} \\ 
  Miyagi & \textBF{0.242} & 0.051 & 0.203 & \textBF{0.087} & Hyogo & 0.638 & 0.213 & \textBF{0.649} & \textBF{0.327} \\ 
  Akita & \textBF{0.269} & -0.126 & 0.231 & \textBF{-0.124} & Nara & 0.296 & -0.013 & \textBF{0.329} & \textBF{-0.013} \\  
  Yamagata & 0.283 & \textBF{-0.015} & \textBF{0.310} & -0.022 & Wakayama & 0.428 & \textBF{-0.051} & \textBF{0.494} & -0.061 \\  
  Fukushima & 0.370 & 0.050 & \textBF{0.404} & \textBF{0.079} & Tottori & 0.215 & -0.073 & \textBF{0.278} & \textBF{-0.073} \\  
  Ibaraki & 0.526 & 0.094 & \textBF{0.559} & \textBF{0.209} & Shimane & 0.413 & -0.006 & \textBF{0.473} & \textBF{0.032} \\  
  Tochigi & 0.398 & \textBF{0.017} & \textBF{0.447} & -0.005 & Okayama & 0.500 & -0.011 & \textBF{0.520} & \textBF{0.050} \\  
  Gunma & 0.330 & 0.044 & \textBF{0.332} & \textBF{0.081} & Hiroshima & 0.624 & 0.164 & \textBF{0.710} & \textBF{0.268} \\ 
  Saitama & 0.477 & 0.114 & \textBF{0.497} & \textBF{0.129} & Yamaguchi & 0.412 & 0.060 & \textBF{0.473} & \textBF{0.129} \\ 
  Chiba & 0.568 & \textBF{0.079} & \textBF{0.597} & -0.006 & Tokushima & 0.410 & 0.086 & \textBF{0.420} & \textBF{0.144} \\ 
  Tokyo & 0.615 & 0.330 & \textBF{0.675} & \textBF{0.430} & Kagawa & 0.276 & 0.022 & \textBF{0.293} & \textBF{0.120} \\ 
  Kanagawa & 0.548 & 0.134 & \textBF{0.589} & \textBF{0.251} & Ehime & 0.329 & 0.099 & \textBF{0.375} & \textBF{0.205} \\ 
  Niigata & 0.530 & 0.172 & \textBF{0.636} & \textBF{0.179} & Kochi & 0.443 & 0.035 & \textBF{0.483} & \textBF{0.090} \\ 
  Toyama & 0.271 & 0.081 & \textBF{0.280} & \textBF{0.085} & Fukuoka & 0.577 & 0.198 & \textBF{0.603} & \textBF{0.332} \\  
  Ishikawa & 0.159 & -0.009 & \textBF{0.232} & \textBF{0.009} & Saga & 0.254 & -0.046 & \textBF{0.294} & \textBF{-0.017} \\ 
  Fukui & 0.317 & -0.088 & \textBF{0.363} & \textBF{-0.057} & Nagasaki & 0.371 & -0.003 & \textBF{0.395} & \textBF{-0.001} \\  
  Yamanashi & 0.339 & \textBF{-0.009} & \textBF{0.352} & -0.061 & Kumamoto & 0.550 & 0.116 & \textBF{0.581} & \textBF{0.213} \\ 
  Nagano & 0.374 & 0.128 & \textBF{0.453} & \textBF{0.151} & Oita & 0.413 & 0.159 & \textBF{0.424} & \textBF{0.246} \\  
  Gifu & 0.446 & 0.146 & \textBF{0.500} & \textBF{0.214} & Miyazaki & 0.377 & 0.113 & \textBF{0.407} & \textBF{0.169} \\ 
  Shizuoka & 0.554 & 0.077 & \textBF{0.629} & \textBF{0.236} & Kagoshima & 0.526 & 0.126 & \textBF{0.580} & \textBF{0.248} \\  
  Aichi & 0.510 & 0.170 & \textBF{0.572} & \textBF{0.219} & Okinawa & 0.488 & 0.123 & \textBF{0.551} & \textBF{0.143} \\ 
  \bottomrule
\end{longtable}
\end{center}
\end{onehalfspace}

\begin{figure}[!htb]
\centering
\includegraphics[width=8.2cm]{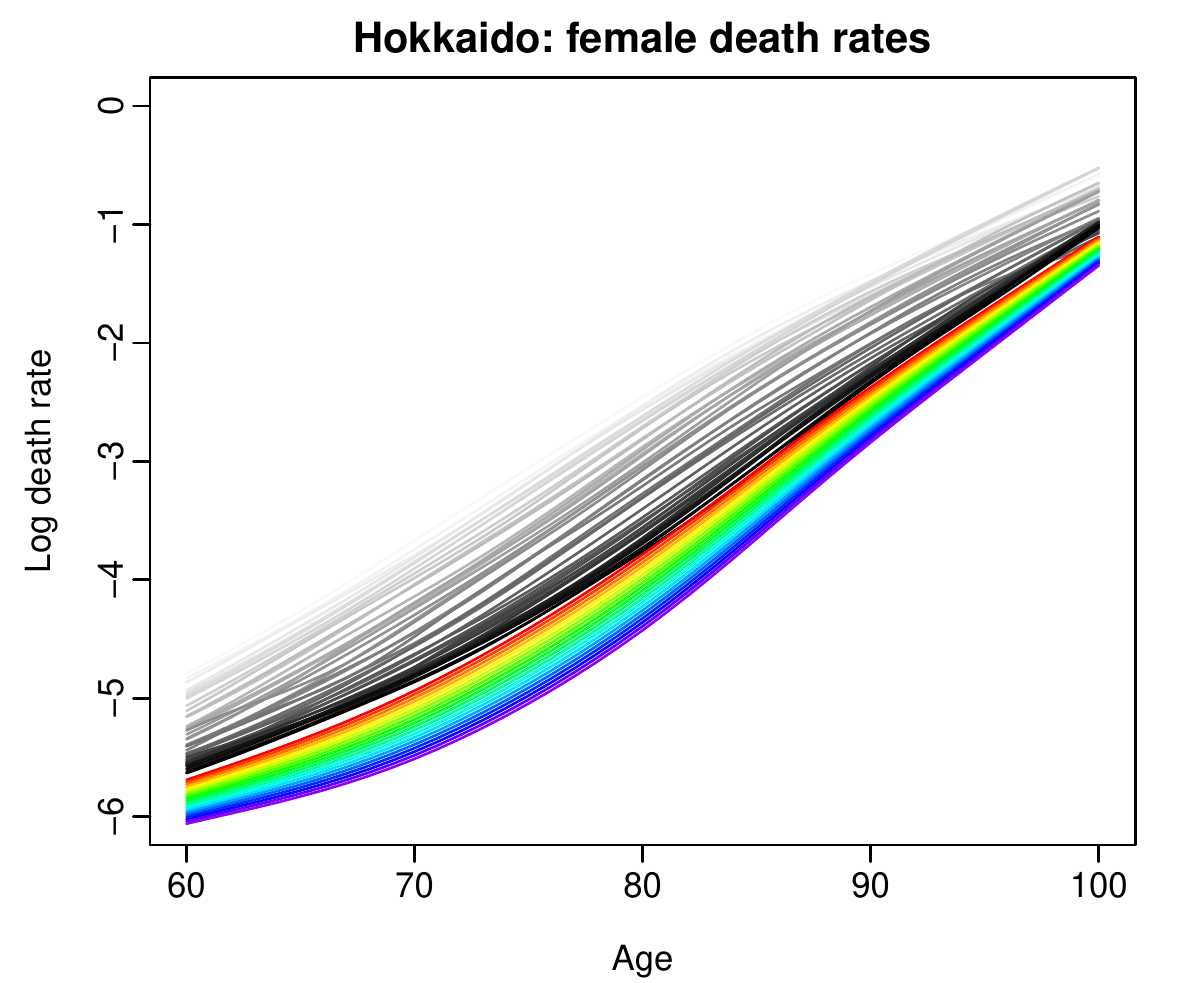}\qquad
\includegraphics[width=8.2cm]{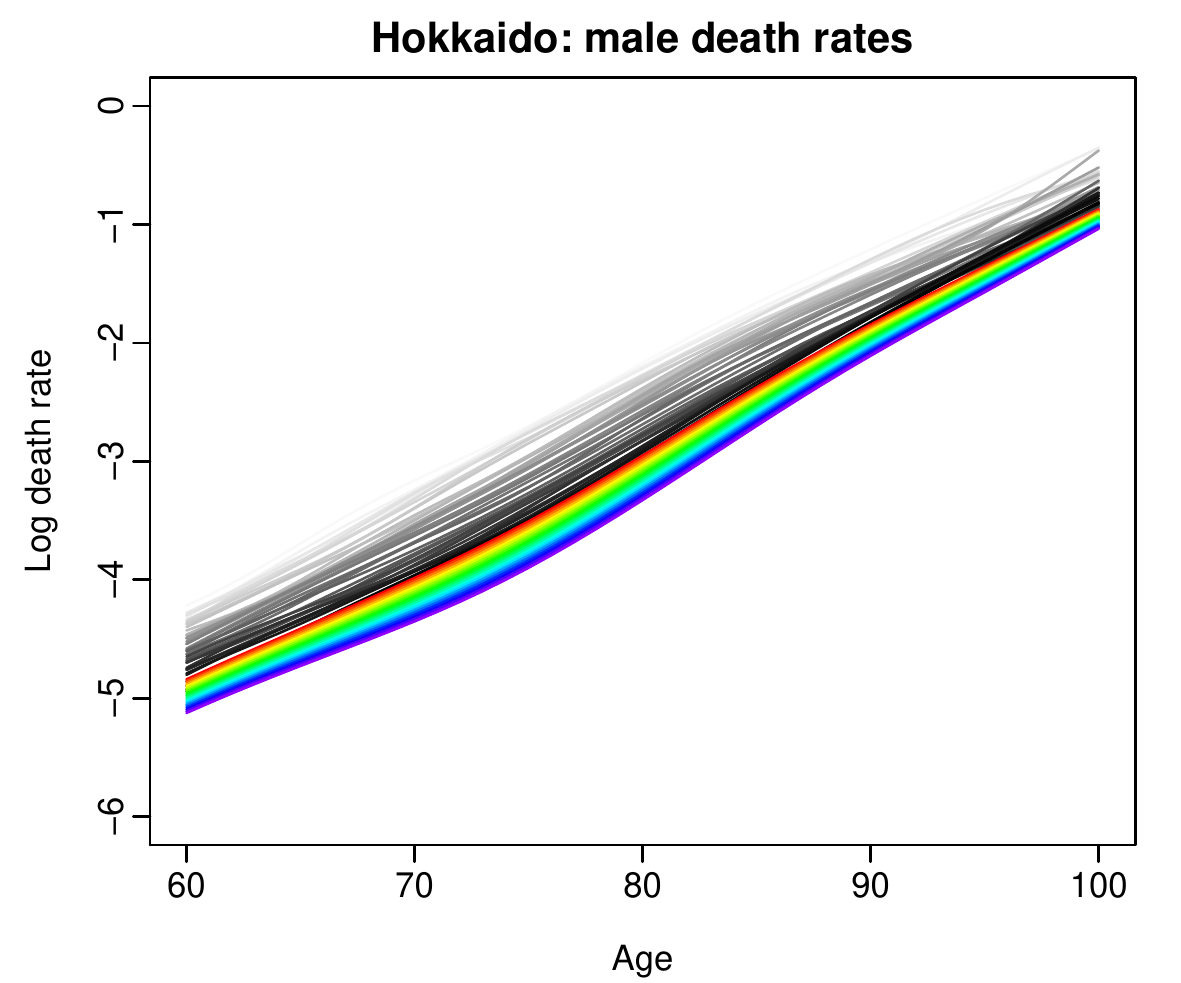}
\caption{Point forecasts of age-specific female and male log mortality rates from 2017 to 2036. The observed functional time series is shown in gray color palette, and the forecasts are highlighted in rainbow color palette.}\label{fig:3}
\end{figure}

Based on the observed mortality rates from 1975 to 2016, we produce the 20-years-ahead point forecasts of female and male age-specific mortality rates from 2017 to 2036. As shown in Figure~\ref{fig:3} for the case of Hokkaido, the age-specific mortality rates are continuing to decline, and the forecast female mortality rates are likely to be lower than the male mortality rates. 

\subsection{Point forecast evaluation}\label{sec:5.1}

Using the first 27 observations from 1975 to 2001 in the Japanese age-specific mortality rates, we produce one- to 15-step-ahead point forecasts. Through an expanding window approach, we re-estimate the parameters in the time series forecasting models using the first 28 observations from 1975 to 2002. Point forecasts from the estimated models are then produced for one- to 14-step-ahead. We repeat this process by increasing the sample size by one year until reaching the end of the data period in 2016. This process produces 15 one-step-ahead forecasts, 14 two-step-ahead forecasts, $\dots$, and one 15-step-ahead forecast. By comparing these forecasts with the holdout samples, we evaluate the out-of-sample point forecast bias and accuracy.

To assess the point forecast bias, we consider the mean forecast error (MFE). For each series $j$, MFE can be written as
\begin{equation}
\text{MFE}_j(h) = \frac{1}{41\times (16-h)}\sum^{15}_{\xi=h}\sum^{41}_{i=1}\left[y_{n+\xi}^{(j)}(x_i) - \widehat{y}^{(j)}_{n+\xi}(x_i)\right],
\end{equation}
where $h$ denotes forecast horizon, $y_{n+\xi}^{(j)}(x_i)$ denotes the actual holdout sample for the $i$\textsuperscript{th} age and $\xi$\textsuperscript{th} curve in the $j$\textsuperscript{th} series, while $\widehat{y}_{n+\xi}^{(j)}(x_i)$ denotes the point forecasts for the holdout sample. To assess the point forecast accuracy, we use the mean absolute forecast error (MAFE) defined as
\begin{equation}
\text{MAFE}_j(h) = \frac{1}{41\times (16-h)}\sum^{15}_{\xi=h}\sum^{41}_{i=1}\left|y_{n+\xi}^{(j)}(x_i) - \widehat{y}^{(j)}_{n+\xi}(x_i)\right|.
\end{equation}

Since the number of series is different across different levels of disaggregation, we obtain an overall assessment of point forecast bias and accuracy by taking the simple average of the error measures across the number of series at each level. With 15 different forecast horizons, we consider the mean and median values to evaluate overall point forecast bias and accuracy between the two functional time-series methods for national and sub-national mortality forecasts.

\subsection{Comparisons of point forecast bias and accuracy}

\begin{figure}[!htbp]
\centering
\includegraphics[width=\textwidth]{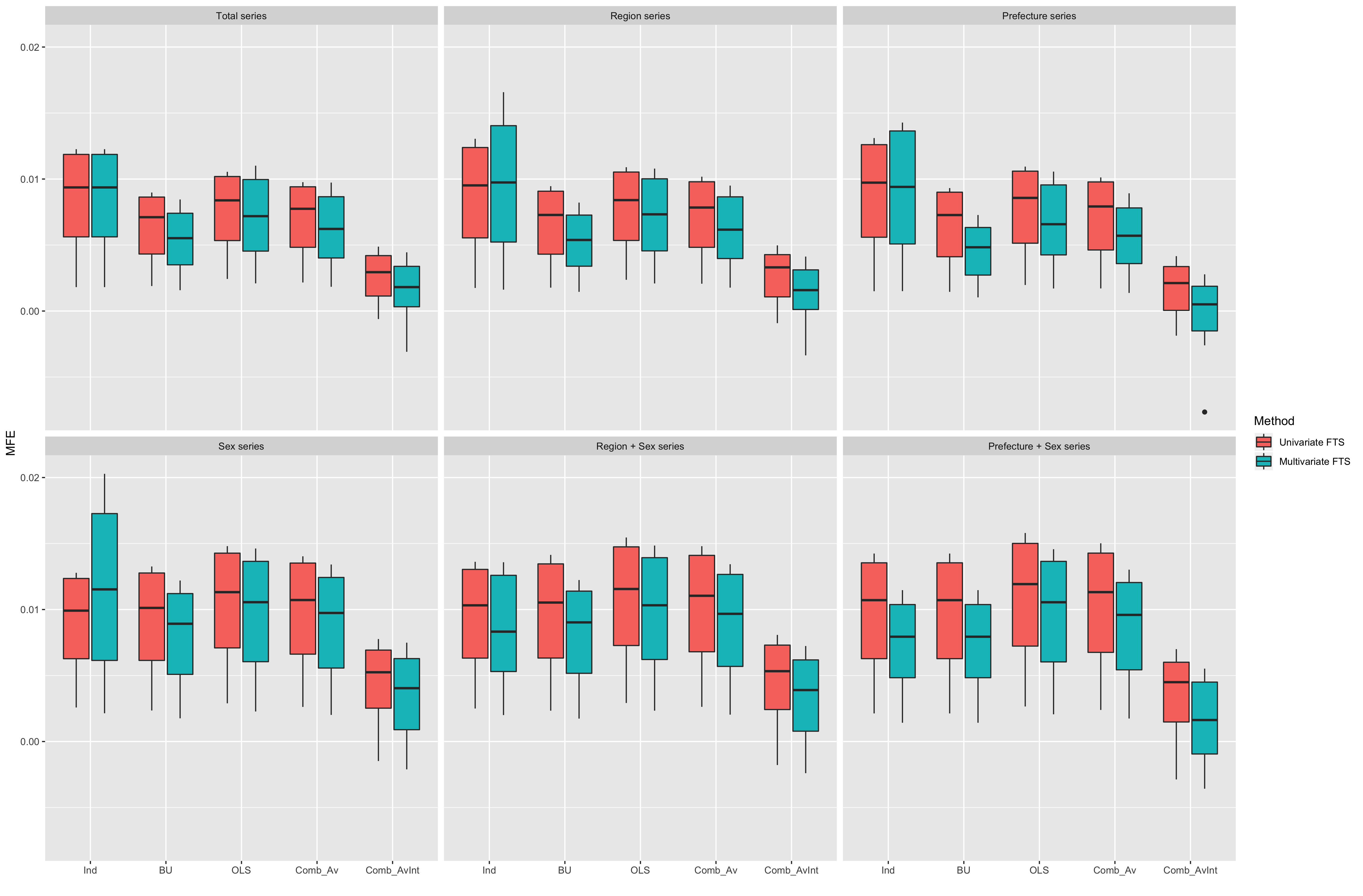}
\caption{MFE in the holdout sample between the univariate and multivariate functional time-series methods applied to the Japanese age-specific mortality rates.\label{fig:MFE_plot}}
\end{figure}

Averaging over all the series at each level of the group structure, Figure~\ref{fig:MFE_plot} compares MFE$(h)$ between the univariate and multivariate functional time-series methods. As measured by the MFE, the multivariate functional time-series method generally produces more accurate point forecast bias than the ones obtained using the univariate functional time-series method. Using the multivariate functional time-series method, the smallest bias can be achieved by using the forecast combination method with simple averaging of midpoints at each level of the group structure. The superior forecast accuracy of the multivariate functional time-series method over the univariate functional time-series method stems from two sources: 
\begin{inparaenum}
\item[(1)] the joint modeling of age-specific mortality patterns among multiple subpopulations; and
\item[(2)] the joint forecasting of age-specific mortality rates among multiple subpopulations.
\end{inparaenum}
The advantage of this forecast combination is that it can reduce bias.

\begin{figure}[!htbp]
\centering
\includegraphics[width=\textwidth]{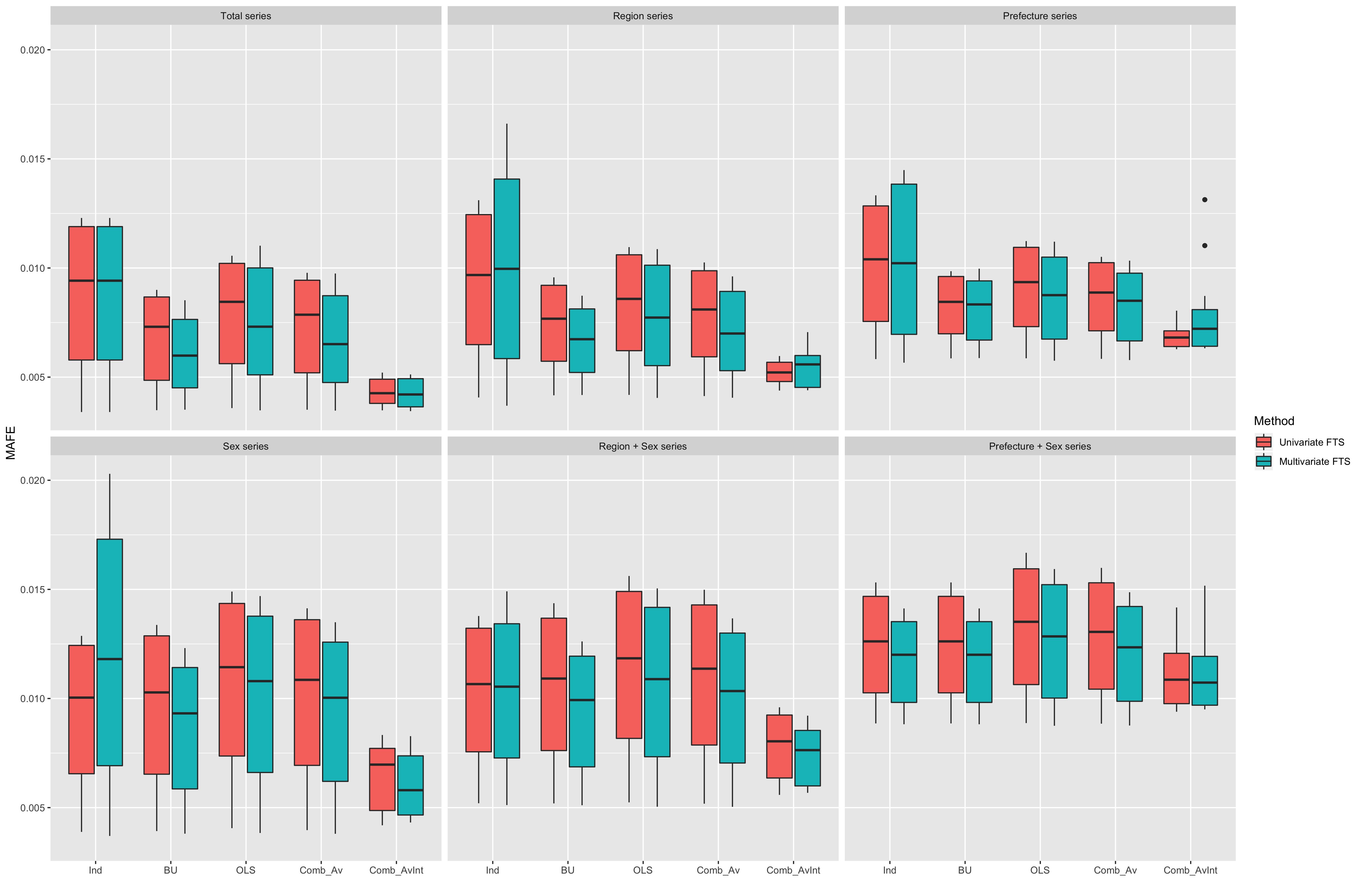}
\caption{MAFE in the holdout sample between the univariate and multivariate functional time-series methods applied to the Japanese age-specific mortality rates.\label{fig:33}}
\end{figure}

Figure~\ref{fig:33} similarly compares MAFE($h$) between the univariate and multivariate functional time-series methods. The multivariate functional time-series method generally produces smaller point forecast errors than the ones obtained using the univariate functional time-series method for almost all levels of the group structure. Based on the two summary statistics of the forecast errors, the multivariate functional time-series method coupled with the forecast combination method (AvInt) performs the best across all levels of the group structure, while it reconciles point forecasts taking account of the group structure. The advantage of this forecast combination is that it can reduce point forecast errors.

\subsection{Interval forecast construction and evaluation}\label{sec:5.2}

We apply the method of \cite{Shang20} to construct pointwise prediction intervals. To evaluate pointwise interval forecast accuracy, we first construct lower and upper bounds of a prediction interval at the $100(1-\alpha)\%$ nominal coverage probability, where $\alpha$ denotes a significance level \citep[see,][for details]{SH17}. Then, we utilize the interval score of \cite{GR07}. In the literature, extensive works are available on interval score and will not be reiterated here. The optimal interval score is achieved when there is almost $100(1-\alpha)\%$ of times that the holdout data lie between the upper and lower bounds of the prediction interval, and the distance between the upper and lower bounds is minimal.

Since the number of series is different across different levels of disaggregation, we obtain an overall assessment of interval forecast accuracy by taking the simple average of the interval scores across the number of series at each level. Also, for 15 different forecast horizons, we consider the mean and median values to evaluate overall interval forecast accuracy among the methods for national and sub-national mortality forecasts.

\subsection{Comparison of interval forecast accuracy}

\begin{figure}[!htbp]
\centering
\includegraphics[width=17.2cm]{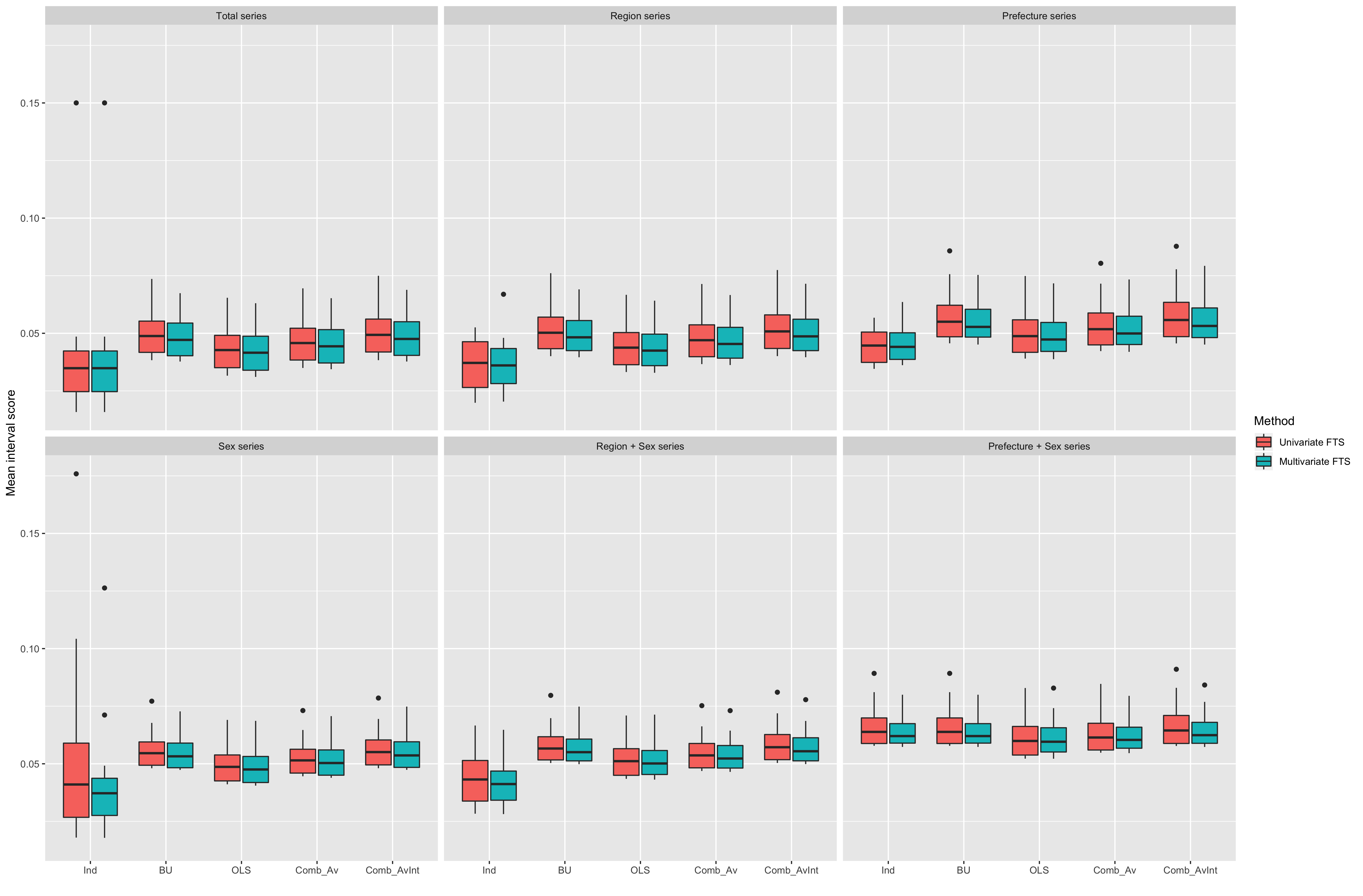}
\caption{Mean interval score in the holdout sample between the univariate and multivariate functional time-series methods applied to the Japanese age-specific mortality rates.}\label{fig:5}
\end{figure}

Averaging over all the series at each level of the group structure, Figure~\ref{fig:5} presents the mean interval scores $\overline{S}_{\alpha}(h)$ between the univariate and multivariate functional time-series methods. Based on the averaged summary statistics of $\overline{S}_{\alpha}(h)$, the independent forecasting method generally performs the best because it fits each series without the constraint of a hierarchy. At the prefecture level, the optimal-combination method outperforms the independent functional time series forecasting method, which demonstrates the improved interval-forecast accuracy of the optimal-combination method while reconciling interval forecasts. Between the univariate and multivariate functional time-series methods, there is a slight advantage to use the multivariate functional time-series method at each level of the group structure. The forecast combination method with equal weighting produces mean interval scores that lie between the ones of the bottom-up and optimal-combination methods. The forecast combination method with simple averaging of midpoints produces the largest mean interval scores and thus this method is conservative.

\section{Application to the pricing of temporary life annuities}\label{sec:6}

An important use of mortality forecasts for elderly is in the pension and insurance industries, whose profitability and solvency rely on accurate mortality forecasts so that longevity risk can be appropriately hedged and pensions and annuities can be accurately valued. When a person reaches retirement age, an optimal way of guaranteeing one individual's financial income in retirement is to purchase an annuity \citep[as demonstrated by][]{Yaari65}. An annuity is a contract offered by insurers guaranteeing a steady stream of payments for the lifetime of the annuitant in exchange for an initial premium fee.  

Lifetime immediate annuities, where rates are locked in for life, have been shown to deliver poor value for money \citep[i.e., they may be expensive for the purchaser: see for example][Chapter 6]{CT08}. In many countries selling annuities, sales of temporary annuities greatly exceed that of lifetime annuities, so this is where a major interest lies in terms of application. These temporary life annuities pay a pre-determined and guaranteed level of income which is often higher than the level of income provided by a lifetime annuity for a similar premium. Temporary annuities offer an alternative to lifetime annuities and allow the purchaser the option of also buying a deferred annuity at a later date. 

We apply the mortality forecasts to the calculation of a temporary life annuity \citep[see][p.114]{DHW09}, and we adopt a cohort approach to the calculation of the survival probabilities. For a single cohort, the $\tau$ year survival probability of a person aged $x$ currently at $t=0$ (or year 2016) is determined by
\begin{align}
_{\tau}p_x &= \prod^{\tau}_{\varpi=1} {}_{1}p_{x+\varpi-1} \\
&= \prod^{\tau}_{\varpi=1}\exp^{-m_{x+\varpi-1, \varpi-1}}.
\end{align}
The survival probability is a random variable given that age-specific mortality rates for $\varpi=1,\dots,\tau$ are forecasts obtained by the multivariate functional time-series method. Here, we assume that the central mortality rates are constant throughout each one-year period \citep[see also][]{SH17}.

The price of a temporary life annuity with a maturity of $T$ years, written for an $x$-year-old with benefit \textyen1 per year and conditional on the path is given by
\begin{align}
a_x^{T}(\bm{m}_{1:T}^x) &= \sum^{T}_{\tau=1}B(0, \tau)\mathbb{E}\left(1_{T_x>\tau}|\bm{m}_{1:\tau}^x\right) \\
&= \sum^T_{\tau=1}B(0,\tau)_{\tau}p_x\left(\bm{m}_{1:\tau}^x\right),
\end{align}
\newpage 

\noindent where $B(0,\tau)$ is the $\tau$-year bond price, $\bm{m}_{1:\tau}^x$ is the first $\tau$ elements of $\bm{m}_{1:T}^x$, and $_{\tau}p_x(\bm{m}_{1:\tau}^x)$ denotes the survival probability given a random $\bm{m}_{1:\tau}^x$ \citep[see also][]{FPS17}. For the purposes of pricing and risk management, it is vital to produce an accurate forecast of the survival curve $_{\tau}p_x$ that best captures the mortality experience of a portfolio.

In Table~\ref{tab:6}, to provide an example of the annuity calculations, we compare the best estimate of the annuity prices for different ages and maturities produced by the three forecasting methods for a female policyholder residing in Region 2. We assume a constant interest rate at $\eta=3\%$ so that the $\tau$-year bond price is given by $B(0,\tau)=\exp^{-\eta \tau}$. 

\begin{onehalfspace}
\begin{center}
  \tabcolsep 0.21in
\begin{longtable}{@{\extracolsep{5pt}} lrrrrrr@{}} 
  \caption{Estimates of temporary life annuity prices with different ages and maturities ($T$) for a female policyholder residing in Region 2. These estimates are based on the forecast age-specific mortality rates from 2017 to 2057, obtained from the independent and grouped multivariate functional time-series forecasting methods. We consider contracts with different ages and maturities, so that age + maturity $\leq 100$.} \label{tab:6} \\
\toprule 
Method & $T=5$ & $T=10$ & $T=15$ & $T=20$ & $T=25$ & $T=30$ \\\midrule
\endfirsthead
Method & $T=5$ & $T=10$ & $T=15$ & $T=20$ & $T=25$ & $T=30$ \\\midrule
\endhead
\hline \multicolumn{7}{r}{{Continued on next page}} 
\endfoot
\endlastfoot
& \multicolumn{6}{c}{\text{age} = 60} \\
\midrule
  Base & 4.5185 & 8.2736 & 11.3152 & 13.6552 & 15.2718 & 16.1779 \\ 
  BU & 4.5192 & 8.2976 & 11.3969 & 13.8402 & 15.6092 & 16.6864 \\ 
  OLS & 4.5244 & 8.3238 & 11.4710 & 14.0044 & 15.9177 & 17.1773 \\ 
  Comb, Av &  4.5218 & 8.3107 & 11.4339 & 13.9218 & 15.7614 & 16.9261 \\
  Comb, AvInt & 4.5272 & 8.3383 & 11.5087 & 14.0769 & 15.8899 & 16.5369 \\
\midrule
& \multicolumn{6}{c}{\text{age} = 65} \\
\midrule
  Base & 4.4672 & 8.0856 & 10.8693 & 12.7925 & 13.8705 & 14.2941 \\ 
  BU & 4.4873 & 8.1681 & 11.0698 & 13.1707 & 14.4500 & 15.0201 \\ 
  OLS & 4.5011 & 8.2297 & 11.2310 & 13.4978 & 14.9900 & 15.7458 \\ 
  Comb, Av & 4.4942 & 8.1988 & 11.1499 & 13.3321 & 14.7136 & 15.3698 \\ 
  Comb, AvInt & 4.5089 & 8.2600 & 11.2984 & 13.4434 & 14.2089 & 14.4340 \\
  \midrule
& \multicolumn{6}{c}{\text{age} = 70} \\
\midrule
  Base & 4.3903 & 7.7677 & 10.1012 & 11.4091 & 11.9231 & 12.0420 \\ 
  BU & 4.4284 & 7.9194 & 10.4470 & 11.9862 & 12.6720 & 12.8564 \\ 
  OLS & 4.4590 & 8.0483 & 10.7592 & 12.5437 & 13.4476 & 13.7413 \\ 
  Comb, Av & 4.4437 & 7.9835 & 10.6010 & 12.2581 & 13.0452 & 13.2779 \\
  Comb, AvInt & 4.4716 & 8.0939 & 10.6509 & 11.5635 & 11.8318 & 11.9313 \\
  \bottomrule
\end{longtable}
\end{center}
\end{onehalfspace}

We highlight that some of the annuity prices are very sensitive to the mortality forecasts obtained from the forecasting method. For example, the annuity prices varies from 11.4091 to 12.5437 for 20-year annuity at age 70. Thus, it is important to compare forecast accuracy of various forecasting methods and provide recommendations as stated in the conclusion.

To measure forecast uncertainty, we obtain the bootstrapped forecasts of the age-specific mortality rates, derive the survival probabilities and calculate the corresponding annuity prices associated with different ages and maturities. For instance, we construct one-step-ahead to 15-step-ahead bootstrapped forecasts of the age-specific mortality rates, derive the bootstrap survival probabilities and calculate the bootstrap prices of temporary life annuities. In Table~\ref{tab:7}, we present the 95\% pointwise prediction intervals of the prices of temporary life annuities for different ages and maturities, where age + maturity $\leq 75$.

\begin{table}[!htbp]
\centering
\tabcolsep 0.46in
\caption{The 95\% pointwise prediction intervals of temporary life annuity prices with different ages and maturities ($T$) for female policyholder residing in Region 2, for example. These estimates are based on the one-step-ahead to 15-step-ahead forecast mortality rates from 2017 to 2031, obtained from the independent and grouped multivariate functional time-series method. We only consider contracts with maturity so that age + maturity $\leq 75$. If age + maturity $>75$, NA will be shown in the table.}\label{tab:7}
\begin{tabular}{@{}lccc@{}}
\toprule
Method & $T=5$ & $T=10$ & $T=15$ \\
\midrule
& \multicolumn{3}{c}{age = 60} \\
\midrule
Base & (4.5214, 4.5343) & (8.3083, 8.3722) & (11.4147, 11.6110) \\ 
BU & (4.5174, 4.5371) & (8.2961, 8.3802) & (11.3923, 11.6242) \\ 
OLS & (4.5192, 4.5360) & (8.3037, 8.3775) & (11.4118, 11.6219) \\ 
Comb, Av & (4.5183, 4.5365) & (8.2999, 8.3789) & (11.4021, 11.6231) \\ 
Comb, En & (4.5169, 4.5370) & (8.2939, 8.3800) & (11.3854, 11.6260) \\
\midrule
& \multicolumn{3}{c}{age = 65} \\
\midrule
Base & (4.4916, 4.5271) & (8.1760, 8.3476) & NA  \\ 
BU & (4.4885, 4.5289) & (8.1662, 8.3518) & NA  \\ 
OLS & (4.4918, 4.5288) & (8.1809, 8.3536) &  NA \\ 
Comb, Av & (4.4901, 4.5288) & (8.1735, 8.3527) &  NA \\ 
Comb, En & (4.4873, 4.5289) & (8.1602, 8.3544) & NA \\
  \midrule
& \multicolumn{3}{c}{age = 70} \\
\midrule
Base & (4.4256, 4.5189) &  NA & NA  \\ 
BU & (4.4226, 4.5191) & NA  & NA  \\ 
OLS & (4.4300, 4.5212) & NA  & NA  \\ 
Comb, Av & (4.4263, 4.5201) & NA  & NA  \\ 
Comb, En & (4.4193, 4.5217) & NA & NA \\
  \bottomrule
\end{tabular}
\end{table}

Accuracy in pricing is not the focus here as we recognize that variations exist in the approaches adopted to pricing in practice. In particular, the assumption made concerning adverse selection and the mortality experience of purchasers of annuities is critical \citep[see][for a fuller discussion]{CT08}. For example, \cite{OB13} assumed, in an Australian context, that for voluntary purchase of annuities, annuitant mortality was assumed to be 30\% lower than population mortality at age 60 and 20\% lower than population mortality from age 90 onwards, with linear interpolation between these ages, based on the advice from life insurance actuarial consultants.

\section{Conclusion}\label{sec:7}

Using the national and sub-national Japanese age-specific mortality rates, we evaluate and compare the point forecast bias and accuracy between the univariate and multivariate functional time-series methods. Based on the forecast bias and accuracy criteria, we show that the proposed multivariate functional time-series method outperforms the univariate functional time-series method used in \cite{SH17}. The superiority of the multivariate functional time-series method is primarily driven by the ability to incorporate correlation among the subpopulations.

We compare the one-step-ahead to 15-step-ahead point forecast accuracy between the independent and the two grouped univariate and multivariate functional time-series forecasting methods. By using the multivariate functional time-series method to produce base forecasts, we consider forecast reconciliation by applying two grouped functional time-series forecasting methods, namely the bottom-up and optimal-combination methods \citep[see also][]{SH17}. Through a forecast combination approach, we consider two weight selection methods. The grouped multivariate functional time-series forecasting methods produce more accurate point forecasts than those obtained by the independent multivariate functional time-series forecasting method, averaged over all levels of the group structure. Also, the grouped multivariate functional time-series forecasting methods produce forecasts that obey the natural group structure, thus giving forecast mortality rates at the sub-national levels that add up to the forecast mortality rates at the national level. Between the two grouped multivariate functional time-series forecasting methods and their forecast combination methods, the forecast combination with simple averaging of midpoints is recommended for producing point forecasts while the optimal-combination method is recommended for producing interval forecasts at the prefecture level for the data that we have considered.

We also apply the independent and the two grouped multivariate functional time-series methods to forecast age-specific mortality rates from 2016 to 2056. We then calculate the cumulative survival probability and obtain the prices of temporary life annuities. As expected, we find that the cumulative survival probability has a pronounced impact on annuity prices. Although temporary life annuity prices do not differ significantly for the mortality forecasts obtained by the four methods, mispricing could have a dramatic effect on a portfolio of annuity contracts. To assess forecast uncertainty, we obtain one-step-ahead to 15-step-ahead forecasts of age-specific mortality rates, derive their survival probabilities and calculate their annuity prices for various ages and maturities.

There are several ways in which this paper can be extended, and we briefly outline seven:
\begin{enumerate}
\item[1)] Subject to the availability of data, the group structure can be disaggregated more finely by considering different causes of death \citep{GS15} or socioeconomic status \citep{VH14}. 
\item[2)] We may consider other multiple population forecasting methods, in particular non-linear forecasting methods, such as neural networks \citep{RW20, RW20b}.
\item[3)] In the Japanese data example, the female and male series are jointly modeled and forecast within each region or each prefecture. The total series are jointly modeled and forecast among regions or prefectures. It may be interesting to explore other combinations for modeling and forecasting multiple subpopulations. 
\item[4)] A weighted least squares method could be used to estimate the regression coefficient in the optimal combination method \citep[see, e.g.,][]{WAH19}. 
\item[5)] In the forecast combination approach, other grouped forecasting methods and choices of weight selections may be considered.
\item[6)] While the functional principal component analysis extracts latent component on the basis of explained variance, we may consider other dimension reduction methods on the basis of autocorrelation, such as the maximum autocorrelation factor and predictive factor decompositions. 
\item[7)] Finally, the methodology can be applied to calculate prices for other types of annuity product, such as the whole-life immediate annuity or deferred annuity.
\end{enumerate}

\section*{Acknowledgments}

The authors are grateful to the comments and suggestions received from the editor, two reviewers and the conference participants at the Fourth International Workshop on Functional and Operatorial Statistics in A Coru\~{n}a, Spain, and seminar participants at the Cass Business School, City, University of London. This research was partially supported by a faculty research grant from the College of Business and Economics at the Australian National University.

\vspace{.2in}
\begin{center}
{\large\bf SUPPLEMENTARY MATERIAL}
\end{center}

\begin{description}
\item[Code for grouped multivariate functional time-series forecasting] The R code to produce point and interval forecasts from the independent and the two grouped univariate and multivariate functional time-series forecasting methods described in the article. (R\_code.R)
\item[Code for Shiny application] The R code to produce a Shiny user interface for plotting every series in the Japanese data group structure. (shiny.zip)
\end{description}

\newpage
\bibliographystyle{agsm}
\bibliography{GMFTS.bib}

\end{document}